 \documentclass[3p,times]{elsarticle}

\usepackage{graphicx}
\usepackage{amsmath}

\usepackage{amssymb}

\begin{document}

\begin{frontmatter}




\title{Reference Distribution Functions for Magnetically Confined Plasmas from the Minimum Entropy Production Theorem and the MaxEnt Principle, subject to the Scale-Invariant Restrictions}


\author{Giorgio Sonnino$^{1*}$, Alessandro Cardinali$^2$, Gyorgy Steinbrecher$^3$, Philippe Peeters$^1$, Alberto Sonnino$^4$, Pasquale Nardone$^1$}

\address{$^{1*}$ Universit{\'e} Libre de Bruxelles (U.L.B.), Department of Physics, Campus de la Plaine Code Postal 231 - Boulevard du Triomphe, 1050 Brussels, Belgium.}

\address{$^2$EURATOM-ENEA Fusion Association, Via E.Fermi 45, C.P. 65 - 00044 Frascati (Rome), Italy.}

\address{$^3$EURATOM-MEdC Fusion Association, Physics Faculty, University of Craiova, Str.A.I.Cuza 13, 200585 Craiova, Romania.}

\address{$4$ Universit{\'e} Catholique de Louvain (UCL), Ecole Polytechnique de Louvain (EPL), Rue Archim$\grave{\rm e}$de, 1 bte L6.11.01, 1348 Louvain-la-Neuve, Belgium.}

\begin{abstract}
We derive the expression of the reference distribution function for magnetically confined plasmas far from the thermodynamic equilibrium. The local equilibrium state is fixed by imposing the minimum entropy production theorem and the maximum entropy (MaxEnt) principle, subject to scale invariance restrictions. After a short time, the plasma reaches a state close to the local equilibrium. This state is referred to as the reference state. The aim of this letter is to determine the reference distribution function (RDF) when the local equilibrium state is defined by the above mentioned principles. We prove that the RDF is the stationary solution of a generic family of stochastic processes corresponding to an universal Landau-type equation with white parametric noise. As an example of application, we consider a simple, fully ionized, magnetically confined plasmas, with auxiliary Ohmic heating. The free parameters are linked to the transport coefficients of the magnetically confined plasmas, by the kinetic theory. 
\end{abstract}

\begin{keyword}
Non-Equilibrium Thermodynamics, Magnetized plasmas, Kinetic Theory.

\noindent $^{1*}$Email : gsonnino@ulb.ac.be
\end{keyword}


 \end{frontmatter}


\section{Introduction}\label{int}
The objective of the work is to provide the ground state - distribution function, indicated with ${\mathcal F}^R$, that will act as a reference distribution function (RDF) for the perturbative calculus which is done in the plasma-gyrokinetic (GK) simulations. The work is a combination of two interesting lines of analytical developments: the mathematical theory of distribution functions under constraints \cite{sonnino1}-\cite{sonnino2} and the mathematical framework for the theoretical description of the thermodynamic processes using differential geometry concepts \cite{sonnino3}-\cite{sonnino4} . It is expected to be a useful contribution in the field of plasma-gyrokinetic simulation, a crucial step for simulating turbulent, magnetically confined plasmas. Indeed, starting from an arbitrary initial state, collisions would tend, if they were alone, to bring the system very quickly to a local stationary state. But the slow processes i.e., the free flow and the electromagnetic processes, prevent the plasma from reaching this state. The result is that, after a short time, the plasma reaches a state {\it close} to the {\it local equilibrium}. This state is referred to as the {\it reference state}. From here on, the distribution function evolves on the slow time scale. Notice that the local equilibrium state (LES) is {\it not} a state of thermodynamic equilibrium, because the latter must be homogeneous and stationary. The aim of the letter is to determine the expression of ${\mathcal F}^R$, for open thermodynamic systems close to a local equilibrium state, by statistical thermodynamics. The RDF is obtained by perturbing the local equilibrium state. The LES is defined by assuming the validity of a minimal number of hypotheses: the minimum entropy production principle (MEP) and the maximum entropy principle (MaxEnt principle) under two scale invariance restrictions. We recall that the MEP establishes that, in the Onsager region, if the matrix of the transport coefficients is symmetric, a thermodynamic system relaxes towards a stable steady-state in such a way that the rate of the entropy production strength, $\sigma$, is negative. The inequality is saturated at the steady-state. ${\mathcal{F}}^{R}$ is determined in three steps. First we consider open thermodynamic systems obeying to Prigogine's statistical thermodynamics. Successively, we define the local equilibrium state defined by assuming the validity of the MEP and the MaxEnt principle. Finally, we link the Prigogine probability distribution function with particle{\`{}}s density distribution function.

\noindent PrigogineÕs statistical thermodynamics theory starts from the probability distribution of finding a state in which the values of the fluctuation of a thermodynamic variable, $\beta_i$, lies between  $\beta_i$ and  $\beta_i+d\beta_i$. This probability distribution is 
\begin{equation}\label{i1}
{\mathcal F}={\mathcal N}_0\exp[-\Delta_I S]
\end{equation}
\noindent where ${\mathcal N}_0$ ensures normalization to unity, and we have introduced the {\it dimensionless} ({\it density of} ) {\it entropy production} $\Delta_I S$ \cite{prigogine}. Suffix $I$ stands for {\it irreversibility}. We suppose that the system is subject to ${\tilde N}$ thermodynamic forces. The thermodynamic forces defined as $X^\kappa={\partial \Delta_IS}/{\partial \beta_\kappa}$, and the thermodynamic flows defined as $J_\kappa={d\beta_\kappa}/{dt}$, are linked each others by the following equations \cite{degroot}
\begin{equation}\label{i2}
 \Delta_I S=\int^\beta_{\beta^{eq.}}d_IS\quad ;\quad \frac{d_IS}{dt}=\sum_{\kappa=1}^{\tilde N}  X^\kappa J_k=\int\sigma d{\bf x}\geq 0
\end{equation}
\noindent with $d{\bf x}$ denoting the {\it spatial volume element} and the integration is over the whole volume occupied by the system. Notice that $d_IS$ is {\it not} an exact differential. 
In the next section, we briefly derive the general expression of the reference distribution function by statistical thermodynamics. The derivation of a family of stochastic differential equations (SDE), admitting the ${\mathcal{F}}^{R}$ as stationary DDF solution, can be found in Sec.~\ref{family}. A concrete example of calculation, concerning fully ionized, magnetically confined plasmas, is illustrated in Sec.~\ref{mcp}. Finally, conclusions are given in Sec.~\ref{conclusions}.
\section{Reference distribution function and definition of the local equilibrium state}\label{densitydf}
\vskip 0.2truecm
\noindent In this section, we derive the form of ${\mathcal{F}}^{R}$ by following a purely thermodynamic approach. As usual, the gyro-kinetic (GK) theory makes often use of an initial distribution function of guiding centers. In the GK simulations, as well as in the GK theory, this initial distribution function is usually taken as a reference DDF if it depends only
on the invariants of motion and it evolves slowly from the local equilibrium
state i.e., in such a way that the guiding centers remain confined for
sufficiently long time. After a short transition time, the state of the
plasma remains close to the reference state, ${\mathcal{F}}^{R}$, which results in a small deviation of the local equilibrium state (LES). The expression of the
coefficients of the ${\mathcal{F}}^{R}$ will be determined in the next
section by kinetic theory. The reference DDF is obtained by perturbing the
local equilibrium state. The procedure reported in Ref.~\cite{sonnino2} refers to an open system subject to ${\tilde N}$ thermodynamic forces with the local equilibrium state determined by the following two conditions.
\noindent 
\begin{description}
\item {\it {\bf i)} The local equilibrium state corresponds to the values of the $N$ Prigogine$'$s type (fluctuating) variables $\beta_i$ (with $N< {\tilde N}$) for which the entropy production tends to reach an extreme}.
\end{description}
\noindent This special class of variables $\beta_i$ will be denoted as $\alpha_i$. Hence, $\alpha_i$ with $i=1,\cdots, N< {\tilde N}$, are the fluctuating variables $\beta_i$ of Prigogine's type. By definition, a fluctuation is of Prigogine 's type if the entropy production is expressed in quadratic form with respect to these fluctuations (for an exact definition of PrigogineÕs fluctuations refer to Refs~\cite{prigogine}, \cite{degroot}). Under this assumption, close to the local equilibrium and around the extreme value $\partial\Delta_IS/\partial{\alpha_\kappa}\mid_{\alpha_1\cdots \alpha_N=0}\ =0$ (with $\kappa=1,\cdots ,N$), the entropy production can be brought into the form
\begin{equation}\label{i4}
-\Delta_IS=g_0(\bar\beta)-\frac{1}{2}\sum_{i,j=1}^Ng_{ij}(\bar\beta)\alpha_i\alpha_j+h.o.t.
\end{equation}
\noindent Here, ${\bar\beta}$ stands for the vector $(\beta_{N+1},\cdots, \beta_{\tilde N})$ and $h.o.t.$ for {\it higher order terms}. Hence, ${\bar\beta}$ denotes the set of fluctuations, which are {\it not} of Prigogine's type. Notice that the general DDF, ${\mathcal F}$, becomes a reference DDF ${\mathcal F}^{R}$ when the expression of entropy production is given by Eq.~(\ref{i4}). The DDF related to the variables $\bar\beta$, at $\alpha_i=0$ (with $i=1\cdots N$), reads
\begin{equation}\label{i5}
\mathcal{P}(\bar\beta)\equiv{\mathcal F}^{R}\mid_{\alpha_1\cdots \alpha_N=0}={\mathcal N}_0\exp[-\Delta_IS\mid_{\alpha_1\cdots \alpha_N=0}]={\mathcal N}_0\exp[g_0(\bar\beta)]
\end{equation}
\noindent $\mathcal{P}(\bar\beta)$ is determined by the following condition.
\begin{description}
\item {\it {\bf ii)} At the extremizing values $\alpha_i=0$ with $i=1,\cdots N$, under the scale invariance restrictions, the system tends to evolve towards the maximal entropy configurations}.
\end{description}
\noindent Notice that in Eq.~(\ref{i4}) coefficients $g_{ij}$ are directly linked to the transport coefficients of the system \cite{sonnino2}. With these coefficients we may form a positive definite matrix, which can be diagonalized, obtaining 
\begin{equation}
-\Delta _{I}S=g_{0}(\bar\beta)-\sum_{i,j=1}^{N}\delta _{ij}c_{i}(\bar\beta)(\zeta _{i}-\zeta _{i}^{0})^{2}+h.o.t.  \label{i6}
\end{equation}
\noindent where $\delta _{ij}c_{i}({\bar\beta})$ is a positive definite matrix and $\delta _{ij}$ denotes Kronecker$'$s delta. Eq.~(\ref{i6}) allows describing the entire process in terms of $N$ \textit{independent} processes linked to the $N$ independent fluctuations $\zeta _{1},\cdots, \zeta_{N}$. The expression of the reference density of
distribution function is now expressed through a set of convenient variables $\{\zeta _{i}\}$ (with $i=1,\cdots , N$) of the type, \textit{degrees of advancement} (for a rigorous definition of these variables see, for example, Ref.~\cite{prigogine1}. See also the footnote 
\footnote{We recall that, by definition, the degrees of advancement variables $\zeta
_{j}$ satisfy the condition $\lim_{\zeta _{j}\rightarrow \zeta _{j}^{0}}\xi
_{i}=0$ \cite{prigogine}.})

\noindent As a concrete example of calculation, we shall analyze magnetically confined plasmas. In the case of an axisymmetric magnetically confined plasma, after having performed the guiding center transformation, four independent variables are used as orbit coordinates \cite{balescu1}. These variables are defined as follows. One of these is the {\it poloidal magnetic flux}, $\psi$, which for simplicity we consider not to be a fluctuating variable. Plasma is then subject to three thermodynamic forces (i.e., ${\tilde N}=3$), linked to the three (fluctuating) variables. One of these latter variables is the {\it particle kinetic energy per unit mass}, $w$, defined as $w=(v_{\parallel}^{2}+v_{\perp }^{2})/2$ with $v_{\parallel }$ denoting the parallel component of particle's velocity (which may actually be parallel or antiparallel to the magnetic field), and $v_{\perp }$ the absolute value of the perpendicular velocity \cite{balescu1}. The remaining two fluctuating variables are the \textit{toroidal angular moment}, $P_{\phi }$, the variable, $\lambda $. These quantities are defined as (for a rigorous definition, see any standard textbook such as, for example, \cite{balescu1}) 
\begin{equation}\label{ddf1}
P_{\phi }=\psi +\frac{B_{0}}{\Omega _{0c}}\frac{Fv_{\parallel }}{\mid B\mid}\equiv\zeta_1\quad ;\quad \lambda\equiv\frac{\mu }{w}=\frac{\sin ^{2}\theta _{P}}{2\mid B\mid }\equiv\zeta_2\quad 
\mathrm{with}\quad \mu =\frac{v_{\perp }^{2}}{2\mid B\mid }
\end{equation}
\noindent Here $\Omega _{0c}$ is the \textit{cyclotron frequency} associated with the magnetic field along the magnetic axis, $B_0$. $\mid B\mid $, $F$ and  $\theta _{P}$ denote the magnetic field intensity, the {\it characteristic of axisymmetric toroidal field} depending on the surface function $\psi$ and the \textit{pitch angle}, respectively. $P_{\phi }$ and $\lambda$ are considered as two Prigogine$'$s variables. Notice that, even though these variables depend on $w$, actually their variations are independent with each other. So $P_\phi$, $\lambda$ and $w$ are three {\it independent} variables \cite{balescu1}. We define our LES according to the conditions {\bf i}) and {\bf ii}) submitted to the two-scale invariant restrictions $\mathrm{E}[w]= {\rm const.} >0$ and $\mathrm{E}[\ln (w)]={\rm const.}$ (where $\mathrm{E}[\ ]$ is the expectation operation). From condition {\bf ii}), we obtain the expression of ${\mathcal{P}}(w)$. We found that ${\mathcal{P}}(w)$ is a {\it gamma distribution function} \cite{sonnino2}, \cite{papoulis}
\begin{equation} \label{ddf3}
{\mathcal{P}}(w/\Theta)={\mathcal{N}}_{0}\Bigl(\frac{w}{\Theta }\Bigr)
^{\gamma -1}\exp [-w/\Theta ]
\end{equation}
\noindent where we have introduced the \textit{scale parameter} $\Theta $ and the \textit{shape parameter} $\gamma $. The motivation for the choice of the two-scale invariant restrictions (i.e., $\mathrm{E}[w]= {\rm const.} >0$ and $\mathrm{E}[\ln (w)]={\rm const.}$) as well as the special mathematical properties of the resulting DDF can be found in Section~(\ref{family}) and in Ref.~\cite{sonnino2}. We indicate with ${\widehat{\Gamma}}$ the space spanned by the variables $(\psi, w, P_\phi, \lambda, \phi, \Phi)$, where $\phi$ and $\Phi$ are the {\it toroidal angle} and the {\it gyro-phase angle}, respectively. In this space, the reference state takes the form $d{\mathcal{\widehat{F}}}^{R}={\mathcal{F}}^{R}d{\widehat{\Gamma}}$ with
\begin{equation}\label{ddf4}
d{\mathcal{\widehat{F}}}^{R}={\mathcal{N}}_{0}\Bigl(\frac{w}{\Theta }\Bigr)
^{\gamma -1}\!\!\!\!\!\!\exp [-w/\Theta ]\exp [-c_{1}(w/\Theta )(P_\phi-P_{\phi0})^2]\exp [-c_{2}(w/\Theta )(\lambda-\lambda_{0})^{2}]\!\mid {\mathcal{J}}\mid \ \!\!d{\widehat{\Gamma}}
\end{equation}
\noindent where the scripts $0$ refer to (local) equilibrium values. The phase space volume element $d\Gamma =d\mathbf{x}d\mathbf{v}$ is linked to the volume element $d{\widehat{\Gamma }}$ by 
\begin{equation}\label{ddf5}
d\Gamma =\mid \!{\mathcal{J}}\!\mid d{\widehat{\Gamma }} 
\end{equation}
\noindent with $\mid \!{\mathcal{J}}\!\mid $ denoting the Jacobian between $ d\Gamma $ and $d{\widehat{\Gamma }}$. We mention that if  we interpret our reference DDF as a time and ensemble average of the physical DDF describing turbulent plasma, then the singularity at $w=0$ for $0<\gamma<1$ can be related to the intermittency \cite{sonnino2}. Notice that at the point with coordinates $(P_\phi,\lambda,w)=(P_{\phi 0},\lambda_0,(\gamma -1)\Theta)$, the system satisfies the principle of maximum entropy and the entropy production reaches its extreme value. Let us now suppose that $c_{1,2}(w/\Theta)$ are narrow coefficients with small deviations from the expectation value. In this situation we may expand coefficients $c_{1}$ and $c_{2}$ up to the leading order in $w/\Theta$. By taking into account that $(P_\phi-P_{\phi0})^2\sim v_\parallel^2$ and $(\lambda-\lambda_0)\sim v_\perp^2/w$, we get
\begin{equation}\label{ddf6}
c_{1}(w/\Theta ) \simeq c_{1}^{(0)}\equiv \Bigl(\frac{1}{\Delta P_{\phi }}\Bigr)
^{2}=const. \qquad ;\qquad
c_{2}(w/\Theta )\simeq c_{2}^{(0)}+c_{2}^{(1)}\frac{w}{\Theta }\equiv\frac{1}{\Delta \lambda _{0}}\Bigl(\frac{\Delta \lambda
_{0}}{\Delta \lambda _{1}}+\frac{w}{\Theta }\Bigr)\geq 0
\end{equation}
\noindent where $\Delta P_{\phi }$, $\Delta \lambda _{0}$ and $\Delta \lambda _{1}$ are constants. Finally, the expression for the density distribution function ${\mathcal{F}}^{R}$ reads 
\begin{equation}\label{ddf7}
{\mathcal{F}}^{R}={\mathcal{N}}_{0}\Bigl(\frac{w}{\Theta }\Bigr)^{\gamma
-1}\!\!\!\!\!\!\exp [-w/\Theta ]\exp \Bigl[-\Bigl(\frac{P_{\phi }-P_{\phi 0}
}{\Delta P_{\phi }}\Bigr)^{2}\Bigr]\exp \Bigl[-\Bigl(\frac{\Delta \lambda
_{0}}{\Delta \lambda _{1}}+\frac{w}{\Theta }\Bigr)\frac{(\lambda -\lambda
_{0})^{2}}{\Delta \lambda _{0}}\Bigr]\mid {\mathcal{J}}\mid 
\end{equation}
\noindent where ${\mathcal{N}}_{0}$ ensures normalization to unity 
\begin{equation}\label{ddf8}
\int_{\widehat{\Omega}}d{\mathcal{\widehat{F}}}^{R}=\int_{{\widehat{\Omega}}}{\mathcal{F}}
^{R}d{\widehat{\Gamma}}=1  
\end{equation}
\noindent with $\widehat\Omega$ denoting the phase space-volume in the $\widehat\Gamma$ space. The presence of the free parameter $c_{2}^{(0)}$ is crucial. Indeed, the absence of $c_{2}^{(0)}$ precludes the
possibility of identifying the DDF, given by Eq.~(\ref{ddf7}), with the one
estimated by the neoclassical theory for collisional magnetically confined plasmas (see,
for example, Ref.~\cite{balescu2}). In addition, it allows describing more
complex physical scenarios such as, for example, the \textit{modified bi-Maxwelian distribution function}. Last and not least, in some
physical circumstances, the presence of $c_{2}^{(0)}$ is essential to ensure
the normalization of the DDF. Thermodynamics has been able to determine the
shape of the DDF, but it is unable to fix the seven parameters $\Theta
,\gamma ,P_{\phi 0},\lambda _{0},\Delta P_{\phi },\Delta \lambda _{0},\Delta
\lambda _{1}$. These coefficients are linked to the sources. For easy reference, we report the main balance equations linking the RDF with the entropy sources (i.e., the flux entropy and the entropy production strength).

\vskip 0.2truecm

\noindent $\bullet$ {\bf The entropy flux equation}
\begin{equation}\label{t6}
-\int_{\mathcal V}  d{\bf v}\ [{\bf v}-{\bf u}^\alpha({\bf x})]{\mathcal F}^{\alpha R}({\bf v},{\bf x})\ln{\mathcal F}^{\alpha R}({\bf v},{\bf x})=\frac{1}{T_\alpha}({\bf J}_{{\mathcal E }_L}-{\bf J}_{{\mathcal E}_{Oh.}})
\end{equation}
\noindent where ${\mathcal V}$ is the velocity-volume in the phase-space. ${\bf u}^\alpha({\bf x})$ and $T_\alpha$ are the mean velocity and the the temperature of species $\alpha$, respectively. Moreover, ${\bf J}_{{\mathcal E }_L}$ and ${\bf J}_{{\mathcal E}_{Oh.}}$ indicate the energy loss flux and the Ohmic energy flux, respectively.
\vskip 0.2truecm
\noindent $\bullet$ {\bf The entropy production equation}
\begin{equation}\label{t7}
\sigma^\alpha=\frac{n_\alpha}{\tau_\alpha}\Delta_IS^\alpha=-\sum_{\beta =e,i}\int_{\mathcal V}  d{\bf v}\ [\ln {\mathcal F}^{\alpha R}({\bf v},{\bf x})]{\mathcal K}^{\alpha\beta}
\end{equation}
\noindent Here $\tau_\alpha$ and $n_\alpha$ are the {\it relaxation time} and the {\it number density} of species $\alpha$, respectively. ${\mathcal K}^{\alpha\beta}$ denotes the collisional operator of species $\alpha$ due to $\beta$.
\vskip 0.2truecm

\noindent $\bullet$ {\bf The number density equation and the equation for the mean velocity}
\begin{equation}\label{t8}
n_\alpha({\bf x})=\int_{\mathcal V} d{\bf v}{\mathcal F}^{\alpha R}({\bf v},{\bf x})\quad ;\quad 
n_\alpha({\bf x}) {\bf u}^\alpha({\bf x})=\int_{\mathcal V}  d{\bf v}\ {\bf v}{\mathcal F}^{\alpha R}({\bf v},{\bf x})
\end{equation}

\noindent $\bullet$ {\bf The equation for temperature}
\begin{equation}\label{t9}
n_\alpha({\bf x}) T_\alpha({\bf x})=\frac{1}{2}m_\alpha\int_{\mathcal V}  d{\bf v}\mid{\bf v}-{\bf u}_\alpha\mid^2{\mathcal F}^{\alpha R}({\bf v},{\bf x})\qquad\quad{\rm with}\qquad \alpha=(e,i)
\end{equation}
\noindent where $m_\alpha$ is the mass particle of specie $\alpha$. 

\begin{figure}[!ht]\centering
\label{FDD_P_l_s}\includegraphics[scale=0.37]{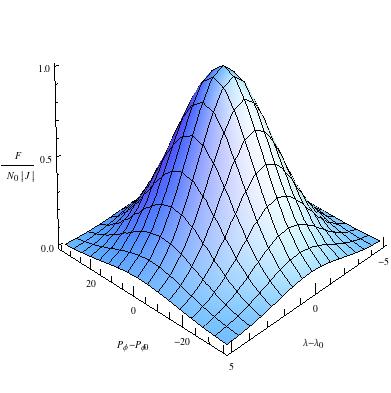}\label{FDD_P_l_c}\qquad\qquad\qquad \includegraphics[scale=0.38]{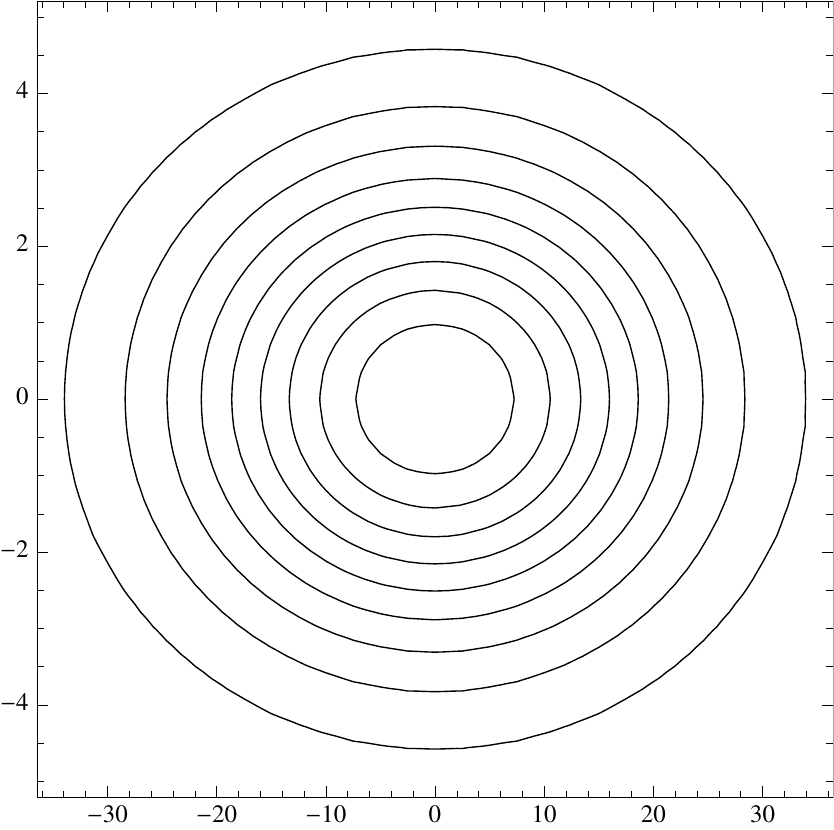} 
\caption{\noindent Distribution function, Eq.~(\ref{ddf7}), computed at  $\gamma=1+E$, $w=E\Theta$, $\Delta P_\phi=22.360$, $\Delta\lambda_0=50.00$ and $\Delta\lambda_1=30.2031$}\label{DDF1}
\end{figure}

\begin{figure}[!ht]\centering
\label{FDD_w_P_s}\includegraphics[scale=0.55]{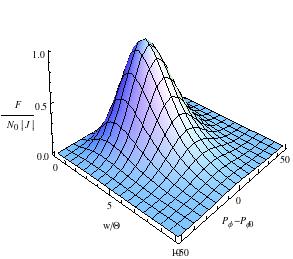}  \label{FDD_w_P_c}\qquad\qquad\qquad\includegraphics[scale=0.54]{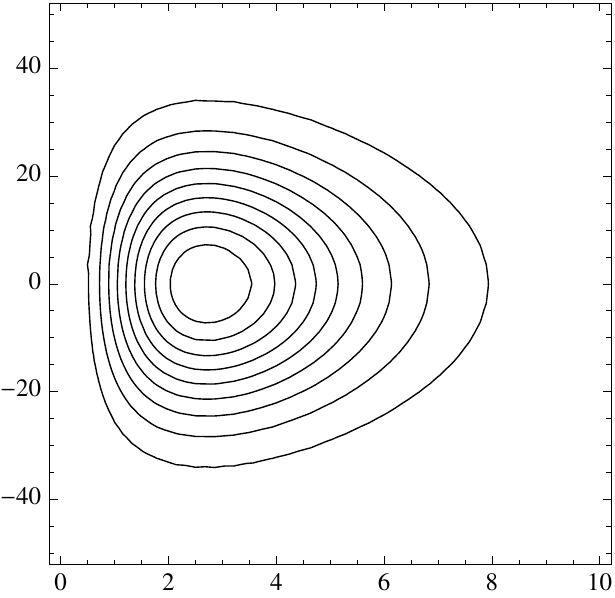} 
\caption{\noindent Distribution function, Eq.~(\ref{ddf7}) computed at $\gamma=1+E$, $\Delta P_\phi=22.360$ and $\lambda=\lambda_0$}\label{DDF2}
\end{figure}

\begin{figure}[!ht]\centering
\label{FDD_w_l_s}\includegraphics[scale=0.55]{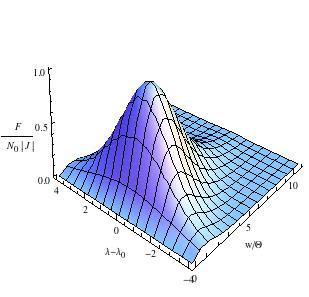} \label{FDD_w_l_c}\qquad\qquad\qquad\includegraphics[scale=0.58]{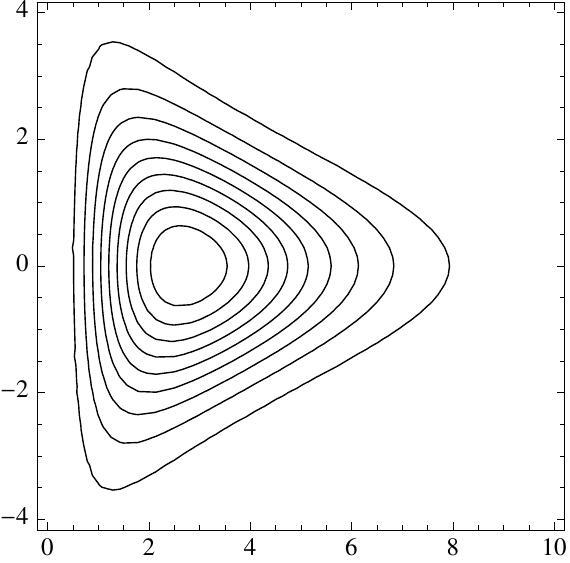} 
\caption{\noindent Distribution function, Eq.~(\ref{ddf7}) computed at $\gamma=1+E$, $P_\phi=P_{\phi0}$, $\Delta\lambda_0=11.111$ and $\Delta\lambda_1=50.00$}. \label{DDF3}
\end{figure}

\noindent Figs~(\ref{DDF1})-(\ref{DDF3}) illustrate three surfaces and contour-plots of Eq.~(\ref{ddf7}) (estimated for unit values of the Jacobian and the normalization coefficient) corresponding to the values $w=E\Theta$ (with $E$ denoting the Euler number), $P_\phi=P_{\phi0}$ and $\lambda=\lambda_0$. 

\section{Generalized class of reference distribution functions subject to scale-invariant restrictions and family of stochastic processes admitting these distribution functions as stationary solutions}\label{family}
\vskip 0.2truecm
\noindent 
Our aim is now to determine the class of stochastic processes whose stationary PDF includes ${\mathcal F}^R$ as a special case. Let us first consider the universal Landau type equation, which includes a multiplicative noise term ${\hat\sigma} dW(t$), with $W$ denoting the standard Brownian motion. In the It${\hat{\rm o}}$ formalism we can write
\begin{equation}\label{inv4}
dw(t)=\bigl({\hat\chi} dt+{\hat\sigma} dW(t)\bigr)w(t)-\varsigma w(t)^2\ dt
\end{equation}
\noindent where ${\hat\chi}>0$ is the {\it instability threshold}, and the function $\varsigma w(t)^2$ is associated with the saturation of the instability controlled by the linear term. This SDE includes the simplest soluble cases of the class of intermittency models \cite{Schenzle}, \cite{Aumaitre}, \cite{Aumaitre2}. The stationary Fokker-Planck equation for the probability density function $\rho(w)$ is \cite{Schenzle}-\cite{Aumaitre}
\begin{equation}\label{inv12}
\frac{\partial}{\partial w}[({\hat\chi}w-\varsigma w^2)\rho(w)]-\frac{{\hat\sigma}^2}{2}\frac{\partial^2}{\partial w^2}[w^2\rho(w)]=0
\end{equation}
\noindent admitting the normalized solution 
\begin{equation}\label{inv5}
\rho(w)=\frac{\varsigma^\gamma}{\Gamma(\gamma)}w^{\gamma-1}\exp[-\varsigma w]\qquad {\rm with}\quad \gamma\equiv\frac{2{\chi}}{{\hat\sigma} ^2}-1>0
\end{equation}
\noindent with $\Gamma(z)$ denoting the Gamma function. It is easily checked that solution (\ref{inv5}) corresponds to Eq.~(\ref{ddf7}) i.e., to the probability density distribution function conditioned by ${\rm E[w] = const.} > 0$ and ${\rm E[ln(w)] = const.}$ 

\noindent Our analysis can be extended to the case where the MaxEnt principle is subject to the general restrictions 
\begin{equation}\label{1.1}
\mathrm{E}[\log(w)] =\nu\quad ;\quad \int_{0}^{\infty}w^{\varsigma_{k}}\rho(w)dw =\mathrm{E}[w^{\varsigma_{k}}]=\mu_k\ \ \ (k=0,1,\cdots,n)
\end{equation}
\noindent where the entropy $S[\rho(.)]$ of probability density function (PDF) $\rho(w)\geq0$ is given by
\begin{equation}
S[\rho(.)]=-\int_{0}^{\infty}\rho(w)\log(\rho(w))dw \label{SG1}
\end{equation}
\noindent Here, $\varsigma_{k},$\ $\mu_{k},\nu$ are real numbers. Observe that the class of restrictions on the PDF given by Eqs~(\ref{1.1}) are invariant under scale transformations. Since we have to consider the important particular cases
\begin{equation}
\int_{0}^{\infty}\rho(w)dw  =1\quad
;\quad \int_{0}^{\infty}w\rho(w)dw =\mathrm{E}[w]=\mu_{1} \label{SG3}
\end{equation}
\noindent it is clear that we must have $n\geq1$ and, in particular, $\varsigma_{0}=0$ ; $\varsigma_1=1$. From Eqs~(\ref{SG3}) we get $\mu_0=1$ and $\mu_1=T/m$, where $T$ and and $m$ are temperature and particle's mass, respectively. By denoting with $\lambda_\kappa$, with $0\leq\kappa\leq n+1$, the Lagrange multipliers in the problem of maximizing the entropy given by Eq.~(\ref{SG1}), with the restrictions (\ref{1.1}), we get
\begin{equation}
\log[\rho(w)] =1-\sum_{\kappa=0}^n\lambda_\kappa w^{\varsigma_\kappa}-\lambda_{n+1}\log(w)\quad\Longrightarrow\quad\rho(w)=C w^{\gamma-1}\exp\Bigl[-\sum_{\kappa=1}^n\lambda_\kappa w^{\varsigma_\kappa}\Bigr]\label{pdf}
\end{equation}
\noindent where $\gamma\equiv1-\lambda_{n+1}$ and $C\equiv\exp(1-\lambda_0)$. Eq.~(\ref{pdf}) provides the generalized class of reference distribution functions subject to the scale-invariant restrictions (\ref{1.1}). Now, our aim is to determine a family of stochastic processes admitting these distribution functions as stationary solutions. Let us consider the following stochastic differential equation (SDE) for the random variable $w(t)$ (which, in our case, corresponds to the energy per unit mass of an individual charged particle). The equation in the It\^{o} form reads 
\begin{equation}
dw(t)=({\hat\chi} dt+{\hat\sigma} dW(t))w(t)-M[w(t)]\ dt\label{z1}
\end{equation}
\noindent where $M[w(t)]$ is the non-linear contribution able to "saturate" the instability triggered by the linear term, with ${\hat \chi}>0$. So we should require that $\underset{w\rightarrow\infty}\lim\frac{M(w)}{w}=+\infty$. As usual, $W(t)$ is the Wiener process and  $\sigma$ is the intensity of the multiplicative noise. Notice that the SDE, Eq.~(\ref{z1}), includes the simplest soluble cases of the class of intermittency models \cite{Schenzle}, \cite{Aumaitre}, \cite{Aumaitre2}. Near $w=0^+$, the solution is dominated by the linear term. So the phenomenology described by Eq.~(\ref{z1}) is still related to the noise-driven intermittency if we require that $\underset{w\rightarrow0^+}{\lim}\frac{M(w)}{w}=0$. As we have seen at the beginning of this section, in the particular case $M(w)=\varsigma w^{2}$, the stationary solution of Eq.~(\ref{z1}) is the gamma distribution. Hence, the class of Eqs~(\ref{z1}) includes the generic type of equations describing the instability growth, on the positive semi-axes (which corresponds to our case), limited by the saturation term. We have slightly modified this equation by adding the random multiplicative noise term ${\hat\sigma} dW(t)$. The stationary Fokker-Planck equation for the density distribution
$\rho(z)$ reads
\begin{equation}\label{z1b}
\frac{\partial}{\partial w}\left[({\hat\chi} w-M(w))\rho(w)\right]
-\frac{{\hat\sigma}^{2}}{2}\frac{\partial^{2}}{\partial w^{2}}\left[  w^{2}
\rho(w)\right]  =0
\end{equation}
\noindent admitting, up to a normalization constant, the following steady state solution
\begin{equation}
\rho(u) =Cw^{\gamma-1}\exp\left[  -\int\frac{M(w)}{w^{2}}dw\right] \quad ;\quad \gamma =\frac{2{\hat\chi}}{{\hat\sigma}^{2}}-1>0\label{z2.1}
\end{equation}
\noindent The general form of $M(w)$, satisfying the conditions for $w\rightarrow 0^+$ and $w\rightarrow\infty$, is
\begin{equation}\label{z4}
M(w) =\sum_{k=1}^{m}A_{k}w^{1+\varkappa_{k}}\qquad {\rm with}\qquad\varkappa_{k} >0
\end{equation}
\noindent with the constraint that, at infinity, the coefficient of the leading term in Eq.(\ref{z4}) should be positive. From Eqs~(\ref{z2.1}) and (\ref{z4}) we obtain
\begin{equation}
\rho(w)=Cw^{\gamma-1}\exp\left[  -\sum_{k=1}^{m}\frac{A_{k}}
{\varkappa_{k}}w^{\varkappa_{k}}\right]  \label{z5}
\end{equation}
\noindent By comparing Eq.~(\ref{z5}) with Eq.~(\ref{pdf}), we can link the exponents $\varsigma_{k}$ used in the restrictions (\ref{1.1}), in the optimization in the MaxEnt principle, with the exponents $\varkappa_{k}$ appearing in the representation of the saturation term in Eq.(\ref{z4}). We find $\varkappa_{k}=\varsigma_{k} $ and $A_{k}=\lambda_{k}\varsigma_{k}$. Notice that, it turns out that, by intermittence mechanism, we obtain the same steady state distribution function with resulting from the MaxEnt principle with the scale invariant restrictions.
\section{Example of calculation. Fully ionized, collisional, magnetically confined plasmas}\label{mcp}
\vskip 0.2truecm
\noindent In this section, we shall give an answer to the following questions:
\begin{description}
\item ${\bullet}$ {\it For collisional magnetically confined plasmas, how much is the deviation $\chi$ of the RDF (${\mathcal{F}}^R$) from the Maxwellian ${\mathcal{F}}^M$, with $\chi$ defined as ${\mathcal{F}}^R={\mathcal{F}}^M(1+\chi)$} ?
\item ${\bullet}$ {\it Does this deviation coincide with the one estimated by the neoclassical theory} (see, for example, Ref.~\cite{balescu2}) ?
\end{description}
\noindent As we shall see, the answer to the latter question is affirmative and, at the same time, such an identification with the neoclassical predictions allows determining the free parameters appearing in the reference RDF, ${\mathcal{F}}^R$.

\noindent To this purpose, let us then consider fully ionized magnetically confined plasmas, defined as a collection of magnetically confined electrons and positively charged ions. In the {\it local triad} (${\bf e}_r,{\bf e}_\theta, {\bf e}_\phi$) (for a rigorous definition refer, for example, to \cite{balescu2}), the magnetic field, in the standard hight aspect ratio, low $beta$ (the plasma pressure normalized to the magnetic field strength), circular tokamak equilibrium model, reads (see, for example, Ref.~\cite{balescu2})
\begin{equation}\label{e1}
{\bf B}=\frac{B_0}{q(r)}\frac{r}{R_0}{\bf e}_\theta+\frac{B_0}{1+(r/R_0)cos\theta}{\bf e}_\phi
\end{equation}
\noindent Here $B_0$ is a constant having the dimension of a magnetic field intensity, and $q(r)$ and $R_0$ are the safety factor and the major radius of the tokamak, respectively. In the magnetic configuration, given by Eq.~(\ref{e1}), we have 
\begin{equation}\label{e2}
\psi(r)=2\pi B_0\int_0^r\frac{r}{q(r)}\ dr
\end{equation}
\noindent According to our formalism, from Eq.~(\ref{i1}) we see that two density distribution functions coincide if, and only if, the entropy productions are identical for {\it all values} taken by the variables. The dimensionless entropy production of species $\alpha$ (with $\alpha=(e,i)$), $\Delta_I S^\alpha$, is derived under the sole assumption that the state of the quiescent plasma is not too far from the reference local Maxwellian. The detailed calculation of these parameters can be found in Ref.~\cite{sonnino2}. In this work we report only the final results and the followed mathematical strategy. In the linear Onsager region, and up to the second order of the drift parameter $\epsilon$, it is possible to show that $\Delta_I S^\alpha$, provided by Eqs~(\ref{i4}) and (\ref{ddf3}), can be brought into the forms \cite{sonnino2}
\begin{equation}\label{e3}
\Delta_IS^e=\frac{1}{2}E\Theta_e^2{X_e^{3}}^2+\frac{1}{2}{\hat g}_{22}^e{X_e^{1}}^2+\frac{1}{2}{\hat g}_{11}^e{X_e^{2}}^2-{\hat g}_{12}^eX_e^{1}X_e^{2}+h.o.t.\quad ;\quad \Delta_IS^i=\frac{1}{2}E\Theta_i^2{X_i^{2}}^2+\frac{1}{2}{\hat g}_{11}^i{X_i^{1}}^2+h.o.t.\nonumber
\end{equation}
\noindent where ${\hat g}^\alpha\equiv g_{ij}^\alpha/g$ (with $g$ indicating the determinant of the matrix $g_{ij}$), and $X_{e,i}^{\kappa}$ (with $\kappa=1,2$) are the electron ($e$) and ion ($i$) thermodynamic forces. Coefficients ${\hat g}^\alpha$ are linked to the transport coefficients by the relations \cite{sonnino2}
\begin{equation}\label{e4}
{\hat g}^e_{11}=\frac{2}{{\tilde\sigma}_\parallel}({\tilde\kappa}^e_\parallel {\tilde\sigma}_\parallel-{\tilde\alpha}_\parallel ^2)\  \ ;\  \ {\hat g}^e_{22}=\frac{2}{{\tilde\sigma}_\parallel}({\tilde\epsilon}_\parallel^e {\tilde\sigma}_\parallel-{\tilde\gamma}_\parallel ^2)\  \ ;\  \ {\hat g}^e_{12}=\frac{2}{{\tilde\sigma}_\parallel}({\tilde\alpha}_\parallel{\tilde\gamma}_\parallel-{\tilde\delta}_\parallel^e {\tilde\sigma}_\parallel)\ \ ;\ \ {\hat g}^i_{11}=\frac{2}{{\tilde\kappa}_\parallel^i}({\tilde\epsilon}_\parallel^i {\tilde\kappa}_\parallel^i-{\tilde\delta}^i_\parallel{}^2)
\end{equation}
\noindent Where ${\tilde{\sigma}}_r$, ${\tilde{\alpha}}_r$, ${\tilde{\kappa}}^{\alpha}_r$ indicate the dimensionless component of the {\it electronic conductivity}, the {\it thermoelectric coefficient} and the {\it electric} ($\alpha =e$) or {\it ion} ($\alpha =i$) {\it thermal conductivity}, respectively. Moreover, ${\tilde\gamma}_\parallel$, ${\tilde\delta}^\alpha_\parallel$ and ${\tilde\epsilon}^\alpha_\parallel$ are the {\it parallel transport coefficients} in 21 Hermitian moment approximation. By using the kinetic equations Eqs~(\ref{t6})-(\ref{t9}), we get the numerical values of the remaining free parameters \cite{sonnino2}
\begin{eqnarray}\label{e5}
&(\Theta_e,\Theta_i)=(4.1760\times 10^{18} cm^2sec^{-2},2.2745\times 10^{15}cm^2sec^{-2})\\
&\Delta {\hat P}_\phi^e= 182.278\qquad;\quad \Delta{\hat\lambda}^e_0\rightarrow\infty\quad ;\quad  \Delta{\hat\lambda}^e_1=581.268\nonumber\\
&(\Delta {\hat P}_\phi^i)^{-2}\sim{\mathcal O}(\epsilon^2)\quad;\quad \Delta{\hat\lambda}^i_0\rightarrow\infty\quad ;\quad  \Delta{\hat\lambda}^i_1=286.236\nonumber
\end{eqnarray}
\noindent where ${\hat P}^\alpha_\phi\equiv P_\alpha/(B_0a^2)$ (with "$a"$ denoting the minor radius of the tokamak), ${\hat\lambda}\equiv B_0\lambda$, and $\epsilon$ denoting the {\it drift parameter}. The electron and ion density distribution functions finally read
\begin{equation}\label{e6a}
{\mathcal F}^{eR}\propto\exp\Bigl[-\Bigl(\frac{{\hat w}-0.5105}{0.3096}\Bigr)^2-\Bigl(\frac{{\hat P}-0.1651}{182.278}\Bigr)^2-\Bigl(\frac{{\hat\lambda}-0.5246}{581.268}\Bigr)^2\ \Bigr]\mid\!{\mathcal J}\!\mid
\end{equation}
\begin{equation}\label{e6b}
{\mathcal F}^{iR}\propto\exp\Bigl[-\Bigl(\frac{10^{3}\times{\hat w} -0.2780}{0.1686}\Bigr)^2-\Bigl(\frac{{\hat\lambda}- 0.3984}{286.236}\Bigr)^2\ \Bigr]\mid\!{\mathcal J}\!\mid
\end{equation}
\noindent where we have introduced the dimensionless variable ${\hat w}\equiv w/v_{the}^2$, with $v_{the}\equiv\sqrt{2T_e/m_e}$ denoting the electron thermal velocity computed at the center of the tokamak. By summarizing, the reference density distribution function ${\mathcal F}^R$, given by Eq.~(\ref{ddf7}), identifies with the reference DDF estimated by the neoclassical theory for collisional magnetically confined plasmas in the Onsager region when the free parameters in Eq.~(\ref{ddf7}) take the values given by Eqs.~(\ref{e5}). In coordinates $\hat w$, ${\hat P}_\phi$ and $\hat\lambda$ (and $\psi$), the expressions of the reference DDFs are given by Eqs~(\ref{e6a}) and (\ref{e6b}). Notice that in this case $c_2^{(1)}=0$ ($\Delta{\hat\lambda}^\alpha_0\rightarrow\infty$) and the presence of the parameter $c_2^{(0)}$ (or of the parameter $\Delta{\hat\lambda}^\alpha_1$) is crucial.
\section{Conclusions}\label{conclusions}
\vskip 0.2truecm
\noindent 

Using statistical thermodynamics approach we have derived the general expression of the (density of) distribution function ${\mathcal F}^R$ for open thermodynamic system where the local equilibrium is fixed by imposing the minimum entropy production theorem and the maximum entropy principle, subject to scale-invariant restrictions. The local equilibrium is fixed by imposing the following conditions.
\begin{description}
\item{{\bf i)} The minimum entropy production theorem is applicable to the fluctuations of Prigogine$'$s type (denoted by $\alpha_i$)};
\item{{\bf ii)} The maximum entropy principle is applicable to the remaining fluctuating variables (denoted by $\beta_{N+1}\cdots\beta_{\tilde N}$)};
\item {{\bf iii)} The scale-invariant restrictions are used in the maximization of the entropy}.
\end{description}

\noindent From this series of ansatzs results a singularity of the RDF that has immediate physical interpretation in terms of the intermittency in turbulent plasmas. The derived RDF, ${\mathcal{F}}^R$, is more general than that currently used for fitting the numerical steady-state solution describing various scenarios of magnetically confined plasmas \cite{pizzuto}-\cite{zonca}. By kinetic theory, we have linked, and then fixed, the seven free parameters entering in ${ \mathcal F}^R$ with the external energy sources and the (internal) entropy production source strength. To be more concrete, we have analyzed the case of, fully ionized, magnetically confined plasmas. This work gives several perspectives. Through the thermodynamical field theory (TFT) \cite{sonnino3} it is possible to estimate the DDF when the nonlinear contributions cannot be neglected \cite{sonnino5}. The next task should be to establish the relation between the reference RDF herein derived with the one found by the TFT. The solution of this difficult problem will contribute to provide a link between a microscopic description and a macroscopic approach (TFT).  Another problem to be solved is the possibility to improve the numerical fit by adding new free parameters according to the principles exposed in this letter.
\section{Acknowledgments}
We thank Dr. Fulvio Zonca, of the Association EURATOM-ENEA, in Frascati (Italy), for having inspired the realization of this work. One of us (G. Sonnino), is very grateful to Prof. M.Malek Mansour, of the Universit{\'e} Libre de Bruxelles, for his scientific suggestions and for his help in the development of this work.



\end{document}